\magnification=\magstep1
\font\bigbfont=cmbx10 scaled\magstep1
\font\bigifont=cmti10 scaled\magstep1

\vsize = 23.5 truecm
\hsize = 15.5 truecm
\hoffset = .2truein
\baselineskip = 14 truept
\overfullrule = 0pt
\parskip = 3 truept
\def\frac#1#2{{#1\over#2}}


\topinsert
\endinsert
\centerline{
{\bigbfont 
KINETIC THEORY AND MESOSCOPIC NOISE}
\footnote*{Based on work presented at the
{\it 23rd International Workshop
on Condensed Matter Theories}, Ithaca, Greece, June 1999.}
}
\vskip 16 truept

\centerline{\bigifont M. P. Das}
\vskip 4 truept
\centerline{
Department of Theoretical Physics}
\centerline{
Research School of Physical Sciences and Engineering}
\centerline{
The Australian National University}
\centerline{
Canberra ACT 0200, Australia}
\vskip 12 truept

\centerline{\bigifont F. Green}
\vskip 4 truept
\centerline{
GaAs IC Prototyping Facility}
\centerline{
CSIRO Telecommunications and Industrial Physics}
\centerline{
PO Box 76, Epping NSW 1710, Australia}

\vskip 1.0 truecm

\centerline{\bf 1. INTRODUCTION} \vskip 12 truept

There are two truisms in the theory of transport. One states that
two-particle correlations carry far more information about
microscopic charge dynamics, than do single-particle processes such as
direct-current response. The other asserts that, for the first insight
to bear fruit, one must look beyond the near-equilibrium limit.

Nowhere are these notions more apt than in testing the
relation between {\it hot-electron noise} and {\it shot noise},
the leading effects of nonequilibrium mesoscopic fluctuations.
Hot-electron noise is generated by spontaneous energy exchanges
between a driven conductor and its thermal bath. Shot noise is
generated by random entry and exit of the discrete carriers.
Neither species is detectable unless a current flows.

Our thesis is that the relationship between shot noise
and hot-electron noise is absolutely fundamental to
understanding mesoscopic fluctuations. At least
from the vantage point of orthodox microscopics and kinetics,
their relation is a long way from being settled.
Its resolution calls for the tools of many-body theory.

In Section 2 we motivate the many-body approach to noise.
Sec. 3 surveys key mesoscopic experiments;
we review the analysis of conductance and noise within
linear diffusive theories, and the physical transition,
or smooth crossover, linking thermal noise and shot noise
[1,2].
In Sec. 4 we outline a kinetic theory of nonequilibrium
fluctuations [3] and discuss how this conventional formulation
directly negates diffusive explanations of the
smooth crossover. We sum up in Sec. 5.

\vskip 28 truept 
\centerline{\bf 2. BACKGROUND}
\vskip 12 truept

At sub-micrometer scales, device sizes approach the
mean free path for scattering and, often, the phase-breaking
length for coherent propagation; they are ``mesoscopic''[4-6],
no longer fitting the usual picture of bulk transport.
Certain structures, such as quantum dots, are tinier still.
Multi-particle correlations are clearly important for devices
supporting only a few carriers at most (strongly quantized dots, say),
but they remain relevant even in a semiclassical setting.
That prompts two questions: What are the experimental signatures
of two-particle correlations at mesoscopic scales? In which ranges
of the driving potential should they be probed?

In dealing with themes similar to the above,
many-body physicists know the value of the van Hove formula.
For solid-state plasma excitations,
it connects the dynamic polarizability
of the electrons directly with inelastic momentum-energy loss,
whenever the system is probed from outside. It forms the
basis of much experimental analysis.
For carrier motion in a conductor, there is a recipe
comparable to van Hove's: the Johnson-Nyquist formula [2].
This connects thermal fluctuations of the current directly
with energy dissipation, both ultimately
induced by the same processes for microscopic scattering.

The connecting principle of the Johnson-Nyquist formula
provides a major consistency criterion for transport models.
Like the van Hove relation, it is an example of
the fluctuation-dissipation theorem [2].
In the electron gas, both share a common basis since
fluctuations, and hence current noise, are microscopically
related to the dynamic polarizability.
Each of the two effects, in its way, reflects the form and action of
the underlying electron-hole excitations.
This drives home a vital, if obvious, message:
{\it a true theory of current noise
cannot avoid being a many-body theory}.

Despite these interconnections, many-body methods are
under-represented in mainstream noise research [7].
With few exceptions [8], the field is
served by special developments of weak-field,
single-particle formalisms. In both quantum coherent and
semiclassical stochastic versions [1,2], the formalisms rest
on novel mesoscopic re-interpretations of drift-diffusion
phenomenology [2,4]. Since noise is
intrinsically a multi-particle effect, the internal logic
of single-particle diffusive approaches
(coherent and stochastic) bears closer inspection [9].

In extending many-body ideas to driven fluctuations,
there are two linked issues:

\item {$\bullet$} Real mesoscopic devices, in real operation,
cannot be characterized by low-field response alone. This is easy to
see in a typical structure 100 nm long, subject to a potential of 0.1 V.
The mean applied field is $10^4 {\rm ~V ~cm}^{-1}$; hardly weak.

\item {$\bullet$} Given the need for a high-field kinetic description,
one must still preserve {\it all} the definitive low-field properties of
the electron gas.

\noindent
While the leading rule of linear transport is the
fluctuation-dissipation theorem (FDT), it is by no means the
sole guiding principle in degenerate Coulomb
systems. The FDT applies within
the context of a nonequilibrium ensemble's adiabatic connection
to the global equilibrium state, whose nature thus
exerts a governing effect on noise.
An electron gas in equilibrium is anything but a collection of
independent carriers. It is a correlated plasma, best known for the
dominance of both degeneracy and quasi-neutrality, which persists
down to distances not far above the Fermi wavelength
and certainly well below mesoscopic [10].

Heuristically, much has been made of the Johnson-Nyquist formula [2]
and the Einstein relation [4], which ties diffusion quantitatively
to conduction in a restricted sense.
However, a model built on these precepts alone is inadequate to
characterize electronic fluctuations [11].
The {\it sum rules} must be respected,
notably compressibility and perfect screening [10].
These ensure quasi-neutrality throughout the degenerate plasma.
Sum rules cannot emerge from an independent-particle analysis
because they refer explicitly to electron-hole correlations.

Theories of fluctuations have less claim to be reliable if they make
many-body predictions by inductive extrapolation from noninteracting
single-particle physics [1,2]. Consider a case in point: the first aim of
all diffusive models is to compute the (one-body) conductance.
To that end, diffusive phenomenology must simply assume
that Einstein's relation and its parent, the microscopic FDT,
are valid [4].
The need to take these theorems on faith removes the logical
possibility of proving them. Without such proof, no
diffusive model can demonstrate its control over the microscopic
multi-particle structure. Without such control, the inner
consistency of any subsequent noise prediction is uncertain.

It is the inability to describe multi-particle correlations {\it ab initio}
that bars access to proof of the fluctuation-dissipation
theorem and the sum rules [9]. All of these constraints will follow
naturally in a canonical description of fluctuations. Conversely,
within a given model of noise in a degenerate conductor,
logical derivation of the constraints is first-hand evidence of
tight control, reliability, and predictive strength.

\vskip 28 truept 
\centerline{\bf 3. LOW-FIELD MESOSCOPIC NOISE}
\vskip 12 truept

There is now a large collection of mesoscopic conductance and noise
measurements. Sample sizes range from a few nanometers,
to hundreds. The experiments generally cover three aspects:
(a) behavior of current noise at low fields,
the only domain in which diffusive theory is valid;
(b) relation of noise to conductance, forming
the tangible link between transport and fluctuations; and
(c) crossover from thermally induced noise to shot noise,
providing a unique signature of the underlying microscopics.
We cite References [1] and [2] for reviews of noise,
and [4], [5], and [6] for mesoscopic transport.

Two simple classes of metallic conductive structures have been tested:
point-contact junctions and diffusive wires [12].
We address each case
separately, and then examine issues of intepretation common to both.

\vskip 12 truept
\centerline{\it Point contacts}
\vskip 8 truept

A point contact is a small conducting region between external leads.
Its aperture approaches the scale of the Fermi wavelength,
narrow enough for transport through it to be
ballistic and strongly quantized [1,4,5]. The contact forms
an ``electron waveguide'' with discrete modes, that is, subbands of
quasi-one-dimensional states propagating through the constriction.
If the junction is fully transmissive, its conductance is
quantized in steps of the universal value ${\cal G}_0 \equiv 2e^2/h$.
This is explained as follows.
Each step signals the opening of a new channel as, with increasing
carrier density, the Fermi surface successively and discretely
intersects higher subbands
(in analogy with the integer quantum-Hall effect [5]).
If the junction has nonideal transmission,
there will be a forward-scattering probability
${\cal T}_n < 1$ for the $n$th mode
at the Fermi energy $\mu$. Each crossing then augments 
the total intrinsic conductance in Landauer's formula

$$
{\cal G} = {\cal G}_0 \sum_n \theta(\varepsilon_n - \mu) {\cal T}_n,
\eqno(1)
$$

\noindent
where $\varepsilon_n$ is the $n$th subband threshold.
There have been many verifications of this result. Two
of the earliest are by van Wees {\it et al.} [13]
and Wharam {\it et al.} [14].

One may ask why, if transport through a point contact
is ballistic (collisionless), its conductance should be finite.
The answer is that the contact is not a closed circuit on its own.
It is open to a larger electrical environment where scattering effects
are strong. The influence of the leads (Landauer's massive banks),
supplying and receiving the current, is paramount [6].
Through dissipative collisions
or by geometrical mismatch at the interfaces, the leads couple
the modes in the contact to an arbitrarily large set of
asymptotic degrees of freedom. This introduces irreversibility
and stabilizes the transport. The details of asymptotic relaxation
should not affect the response; relaxation serves only
to ensure boundedness of the current and the electromotive potential.
In every other way, the relation between them is
an irreducible property of the mesoscopic channel,
albeit in contact with the macroscopic environment [15].

Mesoscopic current-noise measurements are more challenging, owing to
very low signal levels. Good representative data for point contacts
are in Reznikov {\it et al.} [16] and Kumar {\it et al.} [17].
In the zero-temperature limit, there are no thermal
fluctuations; shot noise is the only active form of
carrier correlations. The noise-power spectrum at low frequencies
is [18,19]

$$
{\cal S} = 2eV {\cal G}_0 \sum_n \theta(\varepsilon_n - \mu)
{\cal T}_n(1 - {\cal T}_n)
\eqno(2)
$$ 

\noindent
for voltage $V$ across the contact boundaries.
This theoretical expression is well confirmed by experiment.

To see how ${\cal S}$ relates to the current, take the
single-channel case, for which ${\cal G} = {\cal G}_0 {\cal T}_1$.
Since we are limited to weak voltages,
the current response is linear: $I = {\cal G}V$. We then have

$$
{{\cal S}\over 2eI}  = {{ 2eV {\cal G}_0 {\cal T}_1(1 - {\cal T}_1)}
                  \over {2e{\cal G} V}}
= 1 - {\cal T}_1.
\eqno(3)
$$

\noindent
We have normalized to Schottky's expression [2]
for classical shot noise,
$2eI$, associated with current $I$. 
Equation (3) shows that fluctuations
in the point contact do in fact behave as shot noise,
suppressed below the classical value
depending on transmission. In such a small, quasi-ballistic
device, suppression can {\it only} be a quantum effect.
If the contact is ideal, then ${\cal T}_1 = 1$;
quantum shot noise vanishes completely,
because the incoming and outgoing scattering wave functions overlap
fully with an eigenstate of the system. The asymptotic occupancies
are totally anti-correlated by Pauli exclusion [18,19].
If ${\cal T}_1 < 1$,
the state of the system is no longer asymptotically pure, but mixed.
The occupancies are partly decorrelated,
allowing scope for the appearance of fluctuations.
Evidently,
the fluctuations and their associated shot noise
have a {\it nonlocal} character.

For finite temperatures, with $k_BT \geq eV$, the current noise
displays an appreciable thermal component. In place of Eq. (2),
experimental data [17] follow the expression
(again we keep one channel for simplicity)

$$
{\cal S}(V) = 4k_BT{\cal G}
{\left[
{\cal T}_1 + (1 - {\cal T}_1)
{eV\over 2k_BT} {\rm coth}{\left( {eV\over 2k_BT} \right)}
\right]}.
\eqno(4)
$$

\noindent
This is the prototypical smooth crossover, melding thermal noise
and shot noise into a continuum [19-21].

At equilibrium, Eq. (4) for point-contact noise gives the
classic Johnson-Nyquist form: ${\cal S}(0) = 4k_BT{\cal G}$.
For $eV \gg k_BT$ the second term dominates and
yields quantum shot noise with suppression, just as in Eq. (3).
At intermediate potentials $eV \sim k_BT$,
Eq. (4) takes on a hybrid character,
more than thermal but less than shot.

From Eq. (4) one sees that the suprathermal contribution
${\cal S}(V) - {\cal S}(0)$ has
a quite complex nonlinear dependence on $T$ and $V$.
Eq. (4) is certainly well supported empirically.
In our view, however, the cause of its nonlinearity
is a puzzle in the light of models which depend
(by design) on a strictly {\it linear} drift-diffusion
approach to transport. We revisit this issue shortly.
The smooth crossover also dominates noise in larger
conductors, as we now discuss.

\vskip 12 truept
\centerline{\it Diffusive wires}
\vskip 8 truept

Transport in a diffusive wire is not ballistic, but may still be
quantum-coherent. This is especially so at low temperatures and weak
fields, where scattering is almost perfectly elastic.
If collisions preserve the quantum phase of the carrier wave
function, its total phase shift in transmission depends only
on the total length of the randomized path; this is quantum diffusion.
Samples are too cold, and still too short,
for local dissipative heating.
Instead, carriers thermalize in the access leads [22].

Quantum-mechanically one can think of a diffusive wire as the extreme
limit of a point contact. The subband mode distribution becomes
complicated and quasicontinuous, but Eqs. (1) and (4) still apply.
With a statistical estimate of ${\cal T}_n(E)$
at the Fermi energy $E = \mu$, one can do an ensemble average [1,20]
to get $\sum_n {\cal T}_n^2 \to {2\over 3}\sum_n {\cal T}_n$. 
In the context of multiple modes, Eq. (4) generalizes to

$$
\eqalignno{
{\cal S}(V)
&= 4k_BT{\cal G}_0
\sum_n {\left[
{\cal T}_n^2 + {\cal T}_n(1 - {\cal T}_n)
{eV\over 2k_BT} {\rm coth}{\left( {eV\over 2k_BT} \right)}
\right]} \cr
&\to 4k_BT{\cal G}{\left[ {2\over 3}
+ {1\over 3}{eV\over 2k_BT} {\rm coth}{\left( {eV\over 2k_BT} \right)}
\right]},
&(5) \cr
}
$$

\noindent
presenting the smooth-crossover formula in its
best-known guise [1,2,19-21,23,24].
Once more, the wire at zero voltage exhibits plain Johnson-Nyquist
thermal noise. For $eV \gg k_BT$ there is shot-noise behavior
with the famous threefold suppression; even in conductors
physically much bigger than a point contact, quantum suppression of
shot noise is a robust effect.

As with Eq. (4), there is solid corroboration of Eq. (5) by
experiments [22,25-27]. The fit to measurements is not invariably good.
Besides the survey of boundary-heating effects by
Henny {\it et al.} [22], we note the very early, interesting test of
Eq. (5) by Liefrink {\it et al.} [25] in a two-dimensional electron gas.
That experiment shows clear and systematic departures from
the expected ${1\over 3}$ suppression. Although tentative explanations
have been offered for those deviations, we consider that the
work of Liefrink {\it et al.} in two-dimensional wires has
ongoing importance, and we suggest that it be repeated
with better control over carrier uniformity in the structure [9].

So far, we have reviewed the quantum-coherent interpretation of
diffusive noise theory. Diffusive wires are at the large
end of the mesoscopic range, and elastic scattering
need not {\it necessarily} be phase-preserving. A semiclassical
Boltzmann analysis might be justified if one accepts
that many sequential, locally incoherent collisions should give
much the same diffusive transport as the superposition of
many coherent, but randomly determined, quantum paths.
That is the basis of diffusive adaptations of Boltzmann-Langevin
theory [2].
Such a basis lacks the clarity of the quantum-coherent
descriptions of pure (zero-temperature) shot noise.

This presents an interesting juxtaposition of alternatives:
pure quantum mechanics alongside semiclassical stochastics,
each offering a quite different computational strategy.
We do not retrace the semiclassical derivation here;
theoretical details can be found in the literature [1,2,23,24].
Most important is the fact that these disparate approaches both
converge on Eq. (5). Their agreement, which may seem surprising,
suggests that it is the common assumptions about linear
diffusive transport above all, which matter for the crossover.
If the theoretical crossover were to be disconfirmed,
by whatever means, both derivations would be equally suspect.

\vskip 12 truept
\centerline{\it A Theoretical Issue: Nonlinearity}
\vskip 8 truept

Having already noted the nonlinearity of the crossover formula,
we now examine it more closely. 
Eq. (5), derived either quantum-coherently or
semiclassically, describes all of the fluctuations about a mean
current which is understood to be rigidly linear [1,2,4].
Linearity of the $I$--$V$ relation means that the
resistive power dissipation in the conductor
is strictly quadratic in $V$.

Mesoscopic systems are quite amenable to linear-response analysis
at the microscopic level [3]. If one followed a normal plan for
linear response (such as Kubo's), one would compute a
coefficient (the conductivity) for the local, quadratic, power density.
The calculation, actually a microscopic proof of the FDT, would
furnish the coefficient as a current autocorrelation proportional
to the current-noise spectral density within the conductor.
As an ensemble average at equilibrium, the coefficient could not
depend on the external field, that is, on $V$. 
After integrating it over the sample, the local quantity
would finally lead to ${\cal S}(0)$: Johnson-Nyquist noise,
and nothing more.

In arriving at ${\cal S}(V)$ rather than just ${\cal S}(0)$, diffusive
theories cannot have followed a normal plan for linear response.
Let us run this in reverse. The crossover formula shows marked
dependence of the noise on voltage. On the other hand, it is
derived in a model whose $I$--$V$ response is perfectly linear.
Its power dissipation $IV = {\cal G}V^2$ is perfectly quadratic;
the coefficient ${\cal G}$ must, and does, scale with Johnson-Nyquist
noise as required by the FDT
(naturally so, since the models at hand invoke some
form of Einstein relation, or drift-diffusion FDT, to
secure linearity between $I$ and $V$).
Assuming that the FDT is applicable to any diffusive model
(without benefit of its proof within the same model),
it follows that the excess noise ${\cal S}(V) - {\cal S}(0)$
has no coupling to the equilibrium coefficient fixing the
(strictly quadratic) resistive dissipation.
Thus the excess noise is {\it nondissipative}.

It is evident that the smooth-crossover formula does not fit the
accepted linear-response canon, even though its associated transport
model is in the linear-response regime.
This shows how diffusively based accounts of the
crossover fall short of consistency.
However, it does not touch upon the established
experimental validity of Eq. (5). Indeed, the experiments bring out
one of our themes: the importance of nonequilibrium, nondissipative
noise as a sensitive marker of physical effects on a fine scale
[3].

In terms of theory, two situations arise. Diffusive
models of mesoscopic noise either (i) violate the microscopic
fluctuation-dissipation theorem despite their need to invoke its
offshoot, the Einstein relation, or (ii) they are somehow
covertly nonlinear, despite their manifestly linear construction.
One way or the other, there appear to be problems with
diffusive accounts of the crossover.
Eq. (5) requires a new explanation.

\vskip 28 truept 
\centerline{\bf 4. KINETIC APPROACH}
\vskip 12 truept

We begin by asserting our formalism's most striking conclusion:
{\it the nonequilibrium thermal noise of a
degenerate conductor always scales with bath temperature} $T$.
Since shot noise does not scale with $T$, there is an immediate
corollary. Within kinetic theory,
thermal noise and shot noise cannot be subsumed
under a unified formula.

The focus of this section is on the conceptual structure
of the formalism, with only a brief mathematical overview.
Ref. [3] has more detail.
The kinetic approach to nonequilibrium transport in a metallic
conductor works with a set of assumptions and boundary conditions
identical to those of every other model of current and noise
in metals, including every version of diffusive theory
[1,2,4-6]. They are:

\item
{$\bullet$} an ideal thermal bath regulating the size of
energy exchanges with the conductor, while itself always remaining
in the equilibrium state;

\item
{$\bullet$} ideal macroscopic carrier reservoirs (leads)
in open contact with the conductor, without themselves
being driven out of their local equilibrium;

\item
{$\bullet$} local charge neutrality of the leads,
and overall neutrality of the intervening conductor.

\noindent
This standard scheme, consistently applied within a standard
semiclassical Boltzmann framework, puts tight and explicit
constraints on the behavior of nonequilibrium current noise [3],
constraints that are less transparent in a purely diffusive
framework [9].

The assumption of ideal leads implies that, regardless of the
voltage across the active region, the electron distributions
``far away'' from the conductor remain quiescent and
never depart from their proper equilibrium,
characterized by $T$ and by a uniform density $n$.
In practice, these extended populations need not be further away
than a few Thomas-Fermi screening lengths. The associated
interfacial screening zones will buffer any charge
redistribution; these boundary zones should
be included in the kinetic description of the system.

The electron gas in each asymptotic lead is unconditionally
neutral, and satisfies the canonical sum rules [10].
Gauss' theorem then implies that the central region
must be overall neutral. Global neutrality and asymptotic
equilibrium together condition the form of the
nonequilibrium fluctuations in the mesoscopic conductor.

Our goal is to show that nonequilibrium correlations
are linear functionals of the equilibrium ones.
In the degenerate electron gas, the immediate consequence
of this is that {\it all} thermally induced noise
must scale with ambient temperature $T$.
Therefore it is impossible for shot noise to couple
to the thermal bath. Otherwise, shot noise too
would be seen to scale with $T$, which is not the case.

The kinetic approach to fluctuations, sketched out below,
takes as its input the electron-hole pair excitations
in the equilibrium state.
Fermi-liquid theory shows that these pair correlations
form an {\it essential unit},
always with an internal kinematic coupling.
Generally, they cannot be factorized into two
stochastic components autonomously located,
so to speak, on the single-electron energy shell.
In that respect we do not follow Boltzmann-Langevin
analysis for degenerate electrons [2,28].

It is straightforward to specify the distribution of
free electron-hole fluctuations,
$\Delta f^{\rm eq}_{\bf k}({\bf r})$, for wavevector ${\bf k}$
at position ${\bf r}$:

$$
\Delta f^{\rm eq}_{\bf k}({\bf r}) \equiv
k_BT{{\partial f^{\rm eq}_{\bf k}}
\over {\partial \mu}}
= f^{\rm eq}_{\bf k}({\bf r})[1 - f^{\rm eq}_{\bf k}({\bf r})].
\eqno(6)
$$

\noindent
The one-electron equilibrium distribution is

$$
f^{\rm eq}_{\bf k}({\bf r}) =
{\left[
1 + \exp{\left(
{{\varepsilon_{\bf k} + U_0({\bf r}) - \mu}
\over k_BT}
\right)}
\right]}^{-1},
\eqno(7)
$$

\noindent
where the conduction-band energy $\varepsilon_{\bf k}$
can vary (implicitly) with ${\bf r}$ if the local band
structure varies, as in a heterojunction.
The electronic potential $U_0({\bf r})$ vanishes asymptotically
in the leads, and
satisfies the self-consistent Poisson equation
($\epsilon$ is the background-lattice dielectric constant)

$$
\nabla^2 U_0 \equiv
e{\partial\over {\partial {\bf r}}} \cdot {\bf E}_0
= -{{4 \pi e^2}\over \epsilon}
{\Bigl( \langle f^{\rm eq}({\bf r}) \rangle - n^+({\bf r}) \Bigr)}
\eqno(8)
$$

\noindent
in which, for later use,  ${\bf E}_0({\bf r})$
is the internal field at equilibrium and
${\langle {~~} \rangle}$ denotes the
trace over spin and wave vector ${\bf k}$.
The (nonuniform) neutralizing
background density $n^+({\bf r})$ goes to
$n$ in the (uniform) leads.

The semiclassical Boltzmann equation, subject to
the total internal field ${\bf E}({\bf r}, t)$, can be written as

$$
{\left( { \partial\over {\partial t} }
+ {\cal D}_{\bf k;r}[{\bf E}({\bf r}, t)]
\right)} f_{\bf k}({\bf r}, t)
= -{\cal C}_{\bf k;r}[f].
\eqno(9)
$$

\noindent
Here ${\cal D}_{\bf k;r}[{\bf E}] \equiv
{\bf v}_{\bf k}{\cdot}{ \partial/{\partial {\bf r}} }
- { (e{\bf E}/\hbar})
{\cdot}{ \partial/{\partial {\bf k}} }$ is the convective operator
and ${\cal C}_{\bf k;r}[f]$ is the collision operator, whose kernel
(local in real space) is assumed to
satisfy detailed balance, as usual [1-3]. Even
for single-particle impurity
scattering, of immediate concern,
Pauli blocking of the outgoing scattering states still
means that ${\cal C}$ is nonlinear in the
nonequilibrium solution $f_{\bf k}({\bf r}, t)$.

Since we follow the standard Boltzmann formalism, all
of our results will comply with the conservation laws.
The nonlinear properties of these results will extend as far as the
inbuilt limits of the Boltzmann framework; much further
than if they were restricted to the weak-field domain,
as demanded by the drift-diffusion Ansatz [4].
Moreover, since we rely directly on the whole fluctuation
structure provided by Fermi-liquid theory [10], the sum rules
are incorporated.

Our prescription starts by developing
the steady-state nonequilibrium distribution $f_{\bf k}({\bf r})$
as a mapping of the equilibrium distribution, which satisfies

$$
{\cal D}_{\bf k;r}[{\bf E}_0({\bf r})]
f^{\rm eq}_{\bf k}({\bf r})
= 0
= -{\cal C}_{\bf k;r}[f^{\rm eq}],
\eqno(10)
$$

\noindent
the last equality following by detailed balance.
Subtracting the corresponding sides of Eq. (10)
from both sides of the time-independent
version of Eq. (9), and introducing the difference 
$g_{\bf k}({\bf r}) \equiv
f_{\bf k}({\bf r}) - f^{\rm eq}_{\bf k}({\bf r})$,
we obtain

$$
\eqalignno{
\int \! {}&{} d{\bf r'} \!\! \int \! {{2d{\bf k'}}\over (2\pi)^d} 
{\Bigl( {\cal I}_{\bf kk';rr'}
{\cal D}_{\bf k';r'}[{\bf E}({\bf r'})]
+ {\cal C}'_{\bf kk';kr'}[f] \Bigr)}
g_{\bf k'}({\bf r'}) \cr
&= { {e[{\bf E}({\bf r}) - {\bf E}_0({\bf r})]}\over \hbar } \cdot
{ {\partial f^{\rm eq}_{\bf k}}\over {\partial {\bf k}} }({\bf r})
- {\cal C}''_{\bf k;r}[g].
&(11)
}
$$

\noindent
The unit operator in $d$ dimensions is
${\cal I}_{\bf kk';rr'} \equiv
(2\pi)^d\delta({\bf k} - {\bf k'})\delta({\bf r} - {\bf r'})$,
and the linearized operator ${\cal C}'[f]$ is
the variational derivative
${\cal C}'_{\bf kk';rr'}[f] \equiv
\delta {\cal C}_{\bf k;r}[f]/\delta f_{\bf k'}({\bf r'})$.
Last, ${\cal C}''[g] \equiv
{\cal C}[f] - {\cal C}'[f] \! \cdot \! g$
carries the residual nonlinear contributions.
Global neutrality enforces the important constraint
$\int d{\bf r}{\langle g({\bf r}) \rangle} = 0$.

The leading right-hand term in Eq. (11) is responsible
for the functional dependence of $g$ on the equilibrium
distribution. This is important because dependence
on equilibrium-state properties
carries through to the derived steady-state fluctuations.
The electric-field factor can be written as
${\bf E} - {\bf E}_0 \equiv {\bf E}_{\rm ext} + {\bf E}_{\rm ind}$
where ${\bf E}_{\rm ext}({\bf r})$ is the external
driving field,
and the induced field ${\bf E}_{\rm ind}({\bf r})$
obeys

$$
{\partial\over {\partial {\bf r}}}{\cdot} {\bf E}_{\rm ind}
= -{{4\pi e}\over \epsilon} {\langle g({\bf r}) \rangle}.
\eqno(12)
$$

Now we consider the nonequilibrium fluctuation
$\Delta f_{\bf k}({\bf r}, t)$. It satisfies the
linearized Boltzmann equation [29, 30]

$$
\int \! d{\bf r'} \!\! \int \! {{2d{\bf k'}}\over (2\pi)^d} 
{\left[ {\cal I}_{\bf kk';rr'}
{\left(
{\partial\over {\partial t}}
+ {\cal D}_{\bf k';r'}[{\bf E}({\bf r'})]
\right)}
+ {\cal C}'_{\bf kk';kr'}[f] 
\right]}
\Delta f_{\bf k'}({\bf r'} ,t)
= 0. {~~~}
\eqno(13)
$$

\noindent
Given the temporal and spatial boundary constraints
for this equation (causality and global neutrality),
all of the relevant dynamical
properties of the fluctuating electron gas, notably
its current noise, can be obtained.
Its adiabatic $t \to \infty$ limit,
$\Delta f_{\bf k}({\bf r})$,
represents the average strength of
the spontaneous background fluctuations,
induced in steady state by the ideal thermal bath.
It is one of two essential
components that determine the dynamical fluctuations
(the other is the Green function for the
inhomogeneous form of Eq. (13)).
In particular, $\Delta f_{\bf k}({\bf r})$
dictates the explicit $T$-scaling of all thermal effects
through its functional dependence on
the equilibrium distribution
$\Delta f^{\rm eq}_{\bf k}({\bf r})$.
We now show how this comes about. 

Define the variational derivative
$G_{\bf kk'}({\bf r}, {\bf r'}) \equiv
\delta g_{\bf k}({\bf r})/\delta f^{\rm eq}_{\bf k'}({\bf r'})$.
This is a Green-function-like operator
obeying a steady-state equation obtained from Eq. (11)
by taking variations on both sides [3].
The explicit form of $G$ can be derived from knowledge
of the Green function for Eq. (13).
One can verify that

$$
\Delta f_{\bf k}({\bf r}) =
\Delta f^{\rm eq}_{\bf k}({\bf r})
+
\int \! d{\bf r'} \!\! \int \! {{2d{\bf k'}}\over (2\pi)^d}{~}
G_{\bf kk'}({\bf r}, {\bf r'})
\Delta f^{\rm eq}_{\bf k'}({\bf r'})
\eqno(14)
$$

\noindent
satisfies the steady-state form of Eq. (13) identically.
This establishes the linear relationship between
nonequilibrium and equilibrium thermal fluctuations,
and the need for the former to be proportional to $T$
in a degenerate conductor, since then
$\Delta f^{\rm eq}_{\bf k}({\bf r})
\to k_BT\delta(\varepsilon_{\bf k} + U_0({\bf r)} - \mu)$.

Again, charge neutrality enforces upon Eq. (14) the constraint

$$
\int \! d{\bf r} \!\! \int \! {{2d{\bf k}}\over (2\pi)^d}{~}
G_{\bf kk'}({\bf r}, {\bf r'}) = 0
$$

\noindent
for all ${\bf k'}$ and ${\bf r'}$.
Over volume $\Omega$ of the whole
conductor, including its buffer zones,
this leads to the normalization

$$
\int_{\Omega} d{\bf r} {\langle \Delta f({\bf r}) \rangle}
=
\int_{\Omega} d{\bf r} {\langle \Delta f^{\rm eq}({\bf r}) \rangle}.
\eqno(15)
$$

\noindent
One can compare the strict equality in Eq. (15) with the analogous
situation in any of the diffusive noise formulations
[1,2,18-21,23,24]; diffusive fluctuations do {\it not}
fulfill this most basic of physical constraints.
They do not fulfill it because local equilibrium and neutrality
are not guaranteed, in one lead or more
(depending on where a given model chooses to
locate its ``absolute'' chemical potential $\mu$).
Although those asymptotic conditions are implicitly
respected at the level of one-body transport,
they are no longer respected by fluctuations produced
in the diffusive theories' passage to the two-body level.
Such inconsistency could never arise if the sum rules
for the electron gas [10] were in place and operative.

For semiclassical diffusive models, Eq. (15) restores
conformity of the local $\Delta f_{\bf k}({\bf r})$
with the FDT, at the price of suppression.
If a source of semiclassical suppression does exist,
it is genuinely nonequilibrium and it accords with
global neutrality. Furthermore, there is no compelling
reason to expect that {\it any} semiclassical
description -- including ours -- must recover {\it a priori}
the quantum-coherent result for elastic diffusive wires.

As far as we can see, only a quantum treatment can capture
genuinely {\it nonlocal} physics in mesoscopic systems.
However, we still differ on the separate conceptual
issue of a smooth quantum crossover (as such), based
on drift-diffusion ideas.
We speculate that the S-matrix formalism can work without recourse
to drift-diffusion phenomenology. Its coherent nonlocal nature
gives it a certain numerical robustness against
violations of neutrality, a feature not shared by local theories.

Finally, using the tools that we have outlined, it is possible
not only to display the linear functional dependence of
{\it hot-electron} thermal
noise on $\Delta f^{\rm eq}$, but also to prove
the mesoscopic FDT explicitly for semiclassical noise
in the weak-field limit [3]. Beyond this limit there is a
systematic, nonperturbative way of classifying the appreciable
hot-electron contribution. This type of
excess noise has two features: {\it it is not dissipative,
and it still scales with temperature}.
It is just about impossible for it to ``cross over''
into shot noise, which is indisputably non-thermal [3,9].

\vskip 28 truept 
\centerline{\bf 5. SUMMARY}
\vskip 12 truept

Transport and fluctuations at mesoscopic scales reveal new,
intriguing physics requiring theoretical models beyond the
normal methods for extended, uniform systems. In the mesoscopic
regime, even strongly metallic conductors may become nonuniform and
sharply quantized.
In addition, mesoscopic devices are very likely to operate at high fields.
To date, however, most experimental and theoretical work has
engaged only the low-field linear limit.

There are two theoretical responses to these challenges:
make greater efforts within standard microscopics and kinetics,
or revisit simpler phenomenologies and try to stretch those.
For one-body mesoscopics, notably low-field conductance,
the success of diffusively inspired phenomenologies is
impressive. For many-body effects such as current fluctuations,
diffusive theories have also had considerable success,
as witness the prediction of suppressed mesoscopic shot noise,
whose quantum origin is not in debate.

Regardless of their triumphs and their intuitive appeal, diffusive
phenomenologies have not adduced a microscopic rationale for the
apparent attempt to extrapolate the drift-diffusion Ansatz upward
into the hierarchy of multi-particle correlations. In particular,
diffusively based noise models fail to address the sum rules,
and hence to secure them. The sum rules set fundamental constraints,
whose satisfaction is crucial to the correct
representation of many-body phenomena in the electron gas [7].
Current noise is one such phenomenon. Therefore,
noise predictions based on diffusive arguments are less sure
to be well controlled.

One of diffusive noise theory's key results
is the smooth crossover of equilibrium
thermal noise (scaling with ambient temperature $T$)
into nonequilibrium shot noise (independent of $T$). We have pointed out
that the accepted explanation for the crossover
is incompatible with conventional kinetics and
the theory of charged Fermi liquids.
This is shown by the mandatory $T$-scaling of nonequilibrium thermal
noise in degenerate mesoscopic conductors. Such scaling clearly
precludes any continuous crossover between shot noise
and noise that is generated purely thermally.

In sum, the account of the smooth crossover
given by diffusive analysis is not supported {\it theoretically}
by orthodox kinetic theory.
We believe that its {\it empirical} truth still stands in need
of a more rigorous, and certainly quite different, description.
New experiments would be needed to test any alternative.
We end with a question: since diffusive
analysis itself is widely advertised as a serious
first-principles procedure, could some of
its subsidiary assumptions be defective?

\vskip 28 truept
\centerline{\bf REFERENCES}
\vskip 12 truept

\item{[1]}
M. J. M. de Jong and C. W. J. Beenakker, in
{\it Mesoscopic Electron Transport}, edited by
L. P. Kouwenhoven, G. Sch\"on, and L. L. Sohn,
NATO ASI Series E (Kluwer Academic, Dordrecht, 1997).

\item{[2]}
Sh. M. Kogan, {\it Electronic Noise and Fluctuations in Solids}
(Cambridge University Press, Cambridge, 1996).

\item{[3]}
F. Green and M. P. Das, cond-mat/9809339
(Report RPP3911, CSIRO, unpublished, 1998).

\item{[4]}
S. Datta, {\it Electronic Transport in Mesoscopic Systems}
(Cambridge University Press, Cambridge, 1995).

\item{[5]}
D. K. Ferry and S. M. Goodnick,
{\it Transport in Nanostructures}
(Cambridge University Press, Cambridge, 1997).

\item{[6]}
Y. Imry and R. Landauer, {\it Rev. Mod. Phys.} {\bf 71}, S306 (1999).

\item{[7]}
Many-body issues in mesoscopic noise tend to be presented
as secondary, rather than central, in mainstream thinking.
See for example R. Landauer in
{\it Proceedings of New Phenomena in Mesoscopic Structures, Kauai, 1998}
(submitted to {\it Microelectronic Engineering}).

\item{[8]}
A noteworthy Monte-Carlo study is
P. Tadyszak, F. Danneville, A. Cappy, L. Reggiani,
L. Varani, and L. Rota {\it Appl. Phys. Lett.} {\bf 69} 1450 (1996).

\item{[9]}
F. Green and M. P. Das, in {\it Proceedings of the Second
International Conference on Unsolved Problems of Noise,
Adelaide, 1999} edited by D. Abbott and L. B. Kiss
(AIP, in preparation). See also cond-mat/9905086.

\item{[10]}
D. Pines and P. Nozi\`eres,
{\it The Theory of Quantum Liquids}
(Benjamin, New York, 1966).

\item{[11]}
Indeed the Johnson-Nyquist formula in itself shows
no sensitivity to carrier degeneracy, and neither does the Einstein
relation. See C. Kittel, {\it Elementary Statistical Physics}
(Wiley, New York, 1958), pp 143-5.
Their forms are the same whether the conductor
is classical or degenerate.
Therefore, additional {\it microscopic} input is essential
to any systematic treatment of the fluctuations.

\item{[12]}
We do not discuss noise in a third important,
but more complex, class: tunnel-junction devices [5]. See for example
H. Birk, M. J. M. de Jong, and C. Sch\"onenberger,
{\it Phys. Rev. Lett.} {\bf 75}, 1610 (1995).
There are also remarkable results for tunneling shot noise
in the fractional-quantum-Hall regime. Refer to
R. de Picciotto {\it et al.}  {\it Nature} {\bf 389} 162 (1997);
L. Saminadayar {\it et al.}, {\it Phys. Rev. Lett.} {\bf 79} 2526 (1997);
and M. Reznikov {\it et al.}, {\it Nature} {\bf 399}, 238 (1999).

\item{[13]}
B. J. van Wees, H. van Houten, C. W. J. Beenakker, J. G. Williamson,
L. P. Kouwenhoven, D. van der Marel, and C. T. Foxon,
{\it Phys. Rev. Lett.} {\bf 60}, 848 (1988).

\item{[14]}
D. A. Wharam, T. J. Thornton, R. Newbury, M. Pepper, H. Ahmed, J. E. F. Frost,
D. G. Hasko, D. C. Peacock, D. A. Ritchie, and G. A. C. Jones,
{\it J. Phys. C} {\bf 21}, L209 (1988).
 
\item{[15]}
Landauer's conception extends further, to a perfect
duality between voltage and current by which
{\it either} quantity can equally well induce the other [6].
In conventional kinetic theory, the electromotive force
is always distinguished as the prime cause of the current.
Duality is problematic beyond the weak-field linear limit;
recall the non-monotonic response of a resonant-tunneling diode [5].

\item{[16]}
M. Reznikov,  M. Heiblum, H. Shtrikman, and D. Mahalu,
{\it Phys. Rev. Lett.} {\bf 75} 3340 (1995).

\item{[17]}
A. Kumar, L. Saminadayar, D. C. Glattli, Y. Jin, and B. Etienne,
{\it Phys. Rev. Lett.} {\bf 76} 2778 (1996).

\item{[18]}
V. A. Khlus, {\it Sov. Phys. JETP} {\bf 66}, 1243 (1987);
G. B. Lesovik, {\it JETP Lett.} {\bf 49} 592 (1989).

\item{[19]}
M. B\"uttiker, {\it Phys. Rev. Lett.} {\bf 65} 2901 (1992);
{\it Phys. Rev.} B {\bf 46} 12485 (1992).

\item{[20]}
C. W. J. Beenakker and M. B\"uttiker,
{\it Phys. Rev.} B {\bf 46} 1889 (1992).

\item{[21]}
Th. Martin and R. Landauer, {\it Phys. Rev.} B {\bf 45} 1742 (1992).

\item{[22]}
M. Henny, S. Oberholzer, C. Strunk, and C. Sch\"onenberger,
{\it Phys. Rev. B} {\bf 59}, 2871 (1999).

\item{[23]}
K. E. Nagaev, {\it Phys. Lett.} A {\bf 169} 103 (1992);
{\it Phys. Rev.} B {\bf 52} 4740 (1994).

\item{[24]}
M. J. M. de Jong and C. W. J. Beenakker,
{\it Phys. Rev.} B {\bf 51} 16867 (1995).

\item{[25]}
F. Liefrink, J. I. Dijkhuis, M. J. M. de Jong, L. W. Molenkamp,
and H. van Houten, {\it Phys. Rev.} B {\bf 49} 14066 (1994). 

\item{[26]}
A. H. Steinbach, J. M. Martinis, and M. H. Devoret,
{\it Phys. Rev. Lett.} {\bf 76} 3806 (1996).

\item{[27]}
R. J. Schoelkopf, P. J. Burke, A. A. Kozhevnikov, D. E. Prober,
and M. J. Rooks, {\it Phys. Rev. Lett.} {\bf 78} 3370 (1997).

\item{[28]}
For a critique of Langevin methods in correlated-particle
kinetics, see N. G. van Kampen, {\it Stochastic Processes in Physics
and Chemistry} (North-Holland, Amsterdam, 1981), pp 246-52.

\item{[29]}
S. V. Gantsevich, V. L. Gurevich, and R. Katilius,
{\it Nuovo Cimento} {\bf 2} 1 (1979).

\item{[30]}
Here we freeze the response of the self-consistent
fields. This is equivalent to probing the nonequilibrium
analog of the long-wavelength, ``screened'' Lindhard function [10]
prior to including internal Coulomb screening correlations.
Screening effects are especially
important for inhomogeneous systems [3].
They can be treated systematically in Eq. (13)
in the spirit of a Landau-Silin approach [10].

\end